# Security of Electronic Payment Systems: A Comprehensive Survey

SIAMAK SOLAT, Sorbonne Universités, UPMC University of Paris VI, French National Centre for Scientific Research CNRS, Computer Laboratory

This comprehensive survey deliberated over the security of electronic payment systems. In our research, we focused on either dominant systems or new attempts and innovations to improve the level of security of the electronic payment systems. This survey consists of the Card-present (CP) transactions and a review of its dominant system i.e. EMV including several researches at Cambridge university to designate variant types of attacks against this standard which demonstrates lack of a secure "offline" authentication method that is one of the main purpose of using the smart cards instead of magnetic stripe cards which are not able to participate in authentication process, the evaluation of the EMV migration from RSA cryptosystem to ECC based cryptosystem 3. The evaluation of the Card-not-present transactions approaches including 3D Secure, 3D SET, SET/EMV and EMV/CAP, the impact of concept of Tokenization and the role of Blind Signatures schemes in electronic cash and E-payment systems, use of quantum key distribution (QKD) in electronic payment systems to achieve unconditional security rather than only computational assurance of the security level by using traditional cryptography, the evaluation of Near Field Communication (NFC) and the contactless payment systems such as Google wallet, Android Pay and Apple Pay, the assessment of the electronic currency and peer to peer payment systems such as Bitcoin. The criterion of our survey for the measurement and the judgment about the quality of the security in electronic payment systems was this quote: "The security of a system is only as strong as its weakest link" [1].

Key Words and Phrases: Electronic Payment Systems, Security, Card-Present Transactions, Card-Not-Present (CNP) Transactions, Smartcard, Europay-MasterCard-Visa (EMV), Chip and Pin System, Secure Electronic Transaction (SET), 3D Secure, Chip Authentication Program (CAP), Tokenization, Near Field Communication (NFC), Quantum Key Distribution (QKD), Blind Signatures, Electronic Currency, Bitcoin, Cryptocurrency, Elliptic Curve Cryptography (ECC),

## 1. INTRODUCTION

Nowadays the payment approaches via the internet and the network has been growing at a furious pace. In this way, the variety of the electronic payment systems consists of a numerous types to achieve a strong level of the security. However, in parallel, the attacks procedures and strategies are as advanced as the security solutions. In this comprehensive survey, we tried to mention and determine all attempts in both sides (i.e. attacks strategies and security solutions). In this way, we chose the most dominant electronic payment systems and the most successful attack strategies against them.

The rest of this paper consists of Section 2: Card-present (CP) transactions 2.1: a review of dominant CP transactions system i.e. EMV including several researches at Cambridge university to designate variant types of attacks against this standard including MITM, pre-play and relay attacks which demonstrates lack of a secure "offline" authentication method that is one of the main purpose of using the smart cards instead of magnetic stripe cards which are not able to participate in authentication process, Section 3: the evaluation of the EMV migration from RSA cryptosystem to ECC based cryptosystem, Section 4: the evaluation of the Card-not-present transactions approaches including 3D Secure, 3D SET, SET/EMV and EMV/CAP, Section 5 and 6: the impact of concept of Tokenization and the role of Blind Signatures schemes in electronic cash and E-payment systems, respectively, Section 7: using quantum key distribution (QKD) in electronic payment systems to achieve unconditional security rather than only computational assurance of the security level by using traditional cryptography, Section 8: the assessment of the electronic currency and peer to peer payment systems such as bitcoin, and Section 9: Near Field Communication (NFC) and contactless payment systems along with the evaluation of the related technologies such as Google wallet, Android Pay and Apple Pay.

## 2. CARD PRESENT TRANSACTIONS

The name of Card-present-transaction (CP) is derived from presentation of the card to the merchant by the card holder, at the time of transaction. This means that any online payment via internet even in case of using the card by using a card reader (e.g. SET/EMV or EMV/CAP) are classified as Card-not-present (CNP) transactions, because in such these protocols the card holder does not present the card to the merchant and transaction is not performed via a physical point-of-sell (POS).

In a CP transaction, the card information is sent to the payment gateway and after verification, the gateway will send the transaction authorization to the merchant along with a receipt to the card holder. In such this transaction, in case of fraud by customer, the merchant is not liable of the losses because of the payment gateway authorization. [2]

The dominant system in Card-present-transaction is EMV (sometimes called as Chip and PIN). In following section, we describe briefly this system along with its major vulnerabilities against MITM [6], Pre-play [11] and relay [12] attacks.

### 2.1 EMV: Europay, Mastercard, and Visa

Europay, MasterCard, Visa (abbreviated EMV),in some countries known as "chip (i.e. Smartcard) and pin (i.e. a secret code)", is a complex protocol suite and a set of stipulations to interact between a credit smart card and a payment terminal. EMV cards as an IC card that is used to replace magnetic strip cards, consists of a microprocessor which provides transaction's security and other capabilities that is not possible with magnetic stripe cards [3] and merchant is able to use only one terminal for all card brands [4] [5]. The physical aspect of EMV is based on ISO/IEC 7816 [6].

An EMV transaction contains four major steps [7]. The first step is reading the required data from the card by the terminal to process. For this, terminal asks chip card for required data. The next step is to confirmthe authenticity of the card. For this, there are three approaches called as Card Authentication Methods (CAM): "Static-Data-Authentication" (SDA), "Dynamic-Data-Authentication" (DDA) and "Combined-DDA-with-Application-Cryptogram" (CDA). In the rest of this paper we describe vulnerabilities of all these approaches. Third step is verification of the cardholder via a negotiation between the card and the terminal to choose a possible cardholder verification method. In this way, the data element to keep selected verification method is called as CVM (Cardholder Verification Method). The usual possible approaches to authenticate the cardholder are entering a PIN, cardholder's signature or nothing at all [6]. The last step is to authorize for the transaction. For this, the terminal will confirm that the cardholder has enough balance for current transaction [6].

In rest of this paper, we describe all types of card authentication methods (CAMs) and we mention their vulnerabilities against variant attacks.

*2.1.1. Static Data Authentication.* In EMV, the simplest method to authenticate the card is Static Data Authentication. This approach does not provide protection against message replay [148]. In fact, this approach does not require a chip card and it is possible to do that via a magnetic stripe card without need for a chip card. The reason is lack of performing a public key cryptography method in the card side, because in this approach, there is only a static application data signed by the issuer bank stored into the card [8].

In SDA method, the card consists of a static certificate signed by the issuer bank to demonstrate that its data is legal. Because of using a static certificate, it is obvious to copy it and use it in a counterfeit card including an application which accepts any PIN and thus, this type of cards usually called as Yes-card. [9] This type of attack can be protected by "online" transactions thanks to contacting the bank by the merchant to verify the "Message Authentication Code"of the card. This is performed by a common key between the Smartcard and the issuer bank [9].

*2.1.2. Dynamic Data Authentication.* Contrary to SDA, DDA method requires a chip card which is capable to perform a public key cryptography algorithm. This method needs to generate dynamically a "unique cryptogram" for "each transaction" [8]. In this method, each card has of a secret key to confirm its authenticity and on the other hand each terminal has the VISA or MasterCard's public key, to verify the card's certificate by using a short certificate chain [10]. In accordance with [6] even in case of using DDA approach and offline transaction, it is possible to attack against EMV standard.

*2.1.3. Combined DDA with Application Cryptogram.* Regarding to this fact that one of the main purpose of using smart cards rather than magnetic stripe cards was to be able to perform "secure offline" transactions but SDA and even DDA are yet vulnerable against some attacks in case of offline transaction. EMV since 2000 added a new approach to CAM so-called as Combined DDA with Application Cryptogram (abbreviated CDA) in which similar to DDA, the smart card must be capable to perform RSA cryptography algorithm [5][136]. In fact, the processing of CDA is similar to DDA method but with an additional step in which the card must generate the "second" dynamic signature to confirm that the card which is already authenticated is used for "current transaction authorization" process to ensure security against more sophisticated attacks [8]. In rest of this paper we describe vulnerabilities of both DDA and CDA in case of "offline" transactions where EMV standard suffers lack of a secure offline authentication approach in CAMs.

**2.2 Attackagainst SDA**

As mentioned, SDA approach is vulnerable against a replay attack in which the static certificate is copied and written to a counterfeit card and as a result the card responds "Yes" to any entered code, no matter what PIN has been entered and thus often called "Yes-cards" [6].

In SDA only a symmetric key is used and after PIN verification by the card, the terminal sends to the card the transaction data and then the card calculates a MAC over this data which called as "transaction certificate". It is not feasible to give all terminals this key due to a great risk and possibility of forgery if the key is discovered. Thus, the card can prove its authenticity only if the point-of-saleis "connected to the bank" [10].

In the other word, in case of offline transactions, SDA smart cards are even less secure than magnetic stripe cards, due to no need to know the PIN, since it is mission of the card to verify the PIN and the attacker is able to program a counterfeit card to verify any PIN [10].

**2.3 MITM Attackagainst EMV**

In this section we explain the reason of a MITM attack introduced in [6] against DDA and CDA CAMs in case of offline transactions.

Before anything, we describe briefly card holder verification process in which in case of correct PIN, the card's response is 9000 and otherwise is 63Cx, (x: number of PINsentered by client). But the main weak point and flaw in this process is inability of the terminal to know who has sent the response of the card [6]. This flaw makes plan of a MITM attack. Fig.1 demonstrates this attack.

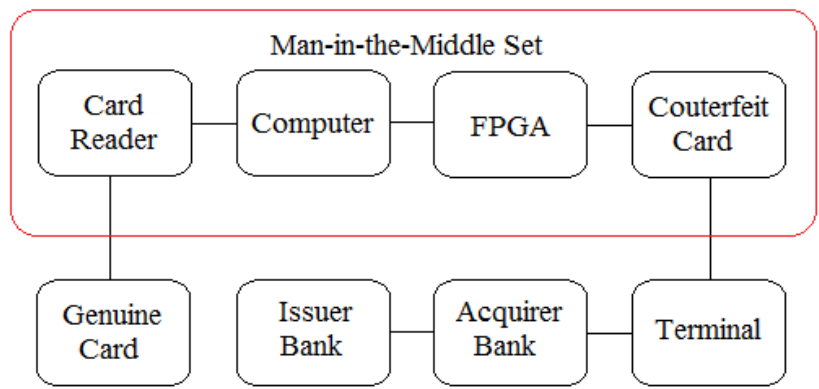

Fig. 1.    MITM attack against EMV

The authenticated data which is sent to the bank contains "Terminal Verification Results" (TVR) and the "Issuer Application Data" (IAD). TVR consists of all possible failed card authentication states which are demonstrated in Table I. But one of the flaws of this process is that in case of successful authentication, it is not mentioned that "which method" has been used (ex. PIN or signature) [6].

Table I. Terminal Verification Results (TVR) In EMV Standard

| bit | Description |
|---|---|
| 1 | Reserved |
| 2 | Reserved |
| 3 | Online PIN entered |
| 4 | PIN is not entered |
| 5 | PIN pad does not work |
| 6 | Number of attempt to enter PIN exceeded |
| 7 | Unrecognized CVM |
| 8 | Cardholder verification was not successful |

*Source*: Terminal Verification Results (TVR) In EMV Standard.

*Note*: TVR consists of all possible failed card authentication states which are demonstrated in Table I. But one of the flaws of this process is that in case of successful authentication, it is not mentioned that "which method" has been used.

Thus, an attacker using this flaw is able to perform a MITM attack to intercept the connection between the card and the point-of-sale to trick the terminal by sending a 9000 response, without transferring the PIN to the genuine card and thereby the genuine card assumes that the point-of-sale does not have "PIN verification" method and uses the signature method to verify the cardholder. On the other hand, because the card has not been received the false PIN, the number of attempts to enter the PIN (i.e. x) is not increased. As a result, the TVR bits are not set, because there are neither attempts nor failure and thereby both the point-of-sale and the card are tricked. The point-of-sale believes that the PIN authentication is successful after receiving 9000 response and then produces a zero value for TVR, thus the card believes the terminal does not support PIN verification method after lack of receiving the PIN and so accepts the terminal's zero byte.

### 2.4 UnpredictableNumber in EMV

In case of the ATM transactions, the chip card is not involved in authentication process, but the ATM encrypts and returns the PIN to the issuer bank. After transaction authorisation via ATM network, the ATM transfers to the card some data including the unpredictable number (abbreviated UN and usually called as "nonce" that is a 32 bit field) to assure the issuer bank that the current transaction is not repeated previously and is fresh. In the response, the card returns the "Authorization Request Cryptogram" (ARQC), which is a MAC calculated over some data including "Application Transaction Counter" (ATC) along with Issuer Application Data (IAD). An attack scenario is feasible by replacing this UN value, introduced in [11] so-called as Pre-Play attack.

The attacker first saves an ARQC using a POS which is infested with a malware. This ARQC is related to nonce N. Then the attacker (in another transaction with an ATM) transfers it to the ATM that generated another nonce i.e. N´. The ATM transfers the ARQC and N´ to the issuer bank (unrelated together). Finally, the attacker by using a MITM device replaces N´ with N and so the transaction is accepted by the issuer bank [11]. Fig.2 demonstrates Pre-play attack.

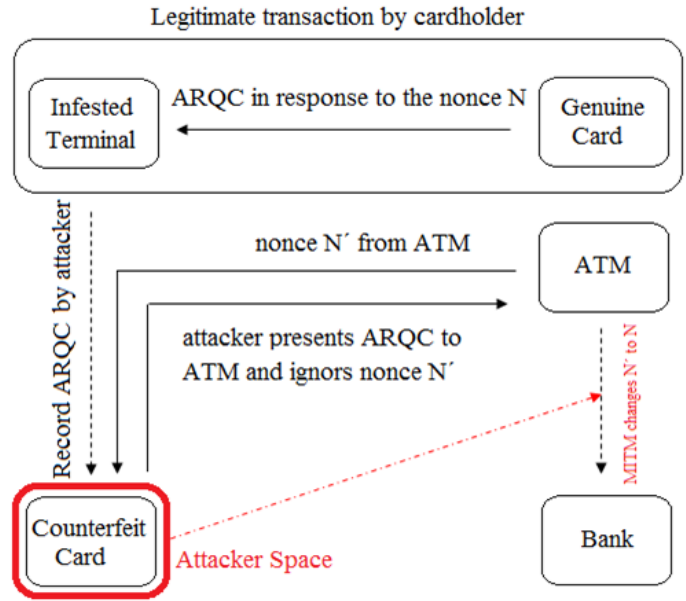

Fig. 2.   Pre-Play attack against EMV

**2.5 Relay Attack against EMV**

Despite the smart card is a tamper resistant device and also convenient to carry, but it suffers a trusted user interface. In the other word, it lacks an input/output to communicate with the cardholder. Lack of an input to enter the PIN directly causes possibility of eavesdropping between keypad and the card. On the other hand, lack of a trusted display leads to inability to confirm the real amount of the transaction and also no information about who is in the other side of the communication which consequently leads to possibility of relay of the data to another place [12].

In accordance with [12], for such this attack, there are some variant approaches to eavesdropping the POS terminal: such as Camera and Double-swipe, Hacked Terminal, Counterfeit Terminal and Terminal Skimmer.

**3. ELLIPTIC CURVE CRYPTOGRAPHY (ECC) IN EMV SMART CARDS**

According to EMV consortium decision, the current RSA based cryptosystem will be replaced by an "Elliptic Curve cryptography system" [149], also the use of both RSA and ECDSA is proposed [150]. A specification for ECC is accessible in [151] which specifies the use of EC-DSA and EC-IES for signatures and encryption, respectively. In another announcement [152], EC-Schnorr is under consideration (which minimizes amount of the computation needed for the signature generation for the smart card that is important due to the limitation in the computational power of the smart cards processors [153]). A main purpose for the migration to ECC is limitation of the current RSA based EMV in size of public keys that is not possible to be larger than 1984 bits [154].

The differences between public key cryptosystems are found in several parameters [150]. The classical public key cryptosystems are based on inequality of performing a calculation against solving the inverse, like RSA (shorted for Rivest, Shamir, Adleman) that is "multiplication"vs."factorisation" [155] or "exponentiation"vs."discrete logarithms" which is used in Diffie-Hellman[156]and Digital Signature Algorithm (DSA)[157].

The RSA is on the basis of *factoring larger integers*, whereas the ECC-based cryptosystem is on the basis of *discrete logarithms* (DL) in the groups of elliptic curve [158].

The fastest known classical efficient factorization algorithm to break RSA is *number field sieve* that runs in *subexponential* time [159] [150] but in case of elliptic curve the fastest algorithm for breaking the "Elliptic Curve Discrete Logarithm Problem" (abbreviate ECDLP) is *"Pollard's rho algorithm"* that runs in *exponential* time and consequently, by using ECDSA, it is possible to achieve the same level of security of RSA but with smaller keys and parameters [150].

Thus, in EMV standard, ECDSA could be an alternative to RSA for reducing both "memory requirement" and "computation power" such that the smart card does not need a co-processor to perform a signing operation. The main difference between RSA and ECDSA signatures is that ECDSA uses a signature with *appendix* (where the signature is usually appended to the unmodified message), whereas RSA signature uses *message recovery* approach (where all or some of the message is embedded in the signature) [150].

In EMV, For the PIN encryption, Elliptic Curve Integrated Encryption Scheme (abbreviated ECIES) could be used that uses Elliptic Curve Diffie-Hellman (abbreviated EC-DH) as a Key Agreement function to generate a key between the card and the terminal along with a symmetric methodlike 3-DES, AES or XOR to encrypt the PIN [150].

The main advantage of ECC based cryptosystems is possibility of providing the security level of RSA but with notably shorter key that is advantageous for the implementation based on the smart card [160].

In accordance with [161], Table II demonstrates size of key in RSA and ECC to provide same level of security. The sizes are in bits. For example, the security level of 80 bits means the best algorithm to break the encrypted messageneedsaround$2^{80}$ steps.

A problem with Public Key Cryptosystems implementation in smart cards is the power of the computation by the smart card processor, where the transaction speed under 100 milliseconds is desired [162] which by using RSA cryptosystem [155] is too long because of numerous operations over large modulus and even in case of using ECC based cryptosystems which is around 10 times faster than RSA [162] and uses smaller key sizes [163] [164], is not still fast enough.

Table II. The security level of 80 bits means the best algorithm to break the encrypted message needs around$2^{80}$ steps

| Security level (bit) | 80 | 112 | 128 | 192 | 256 |
|---|---|---|---|---|---|
| ECC Key Size (bit) | 160 | 224 | 256 | 384 | 512 |
| RSA Key Size (bit) | 1024 | 2048 | 3072 | 8192 | 15360 |

**4. CARD-NOT-PRESENT TRANSACTIONS**

CNP (as the name describes) implies on transactions without presenting the card to the merchant and usually performed via internet or telephone. Because of lack of a physical POS terminal in this type of transactions, there is no standard process to authenticate the cardholder [14] and this is the main challenge issue concerning the security of CNP transaction.

Despite all of weaknesses of security issues in CNP transactions, the CNP fraud is only "one" of the several fraud types. The other frauds are classified as: "First-party fraud", "Counterfeit fraud", "Lost and stolen card fraud", "Mail and non-receipt fraud" and "ID theft" [13].

CNP is divided into two major approaches: The first approach is EMV based protocols which customer at time of transaction need to have his card. The authentication of this type of protocols so-called Two-factor authentication i.e. smart card and password such as SET/EMV and EMV/CAP. In the second approach the customer does not need to have his card at time of transaction, such as 3D SET and 3D SSL (commonly called as 3D Secure and originally branded as "Verified-by-Visa" or "MasterCard-Secure-Code"). In the rest of paper, we describe these protocols.

**4.13D Secure ("Verified-by-Visa" or "MasterCard-Secure-Code")**

The banks largely are interesting to start online authentications of the cards by using the '3D Secure' protocol [125]. In accordance with [125] 3D Secure despite poor technology, is interested by the banks and the merchants.

CNP transactions are performed via the internet, phone or the post in absence of a physical POS terminal in the same location of the card and cardholder; thereby it leads to be a large proportion of the bank losses. Despite reducing frauds by counterfeit cards by replacing the smart cards instead of magnetic stripe cards, but the CNP frauds has been increased significantly [125]. The countermeasure of the industry to this type of frauds is 3D Secure [126]. In the rest of this paper we overview this protocol.

*4.1.13D Secure Overview.* In our review, we focus on Verified by Visa named Visa 3D Secure. The Visa 3D Secure structure is a centralized model including the Directory Servers (DS) which is central point to send the cardholder information between the issuer bank server (Access Control Server (ACS)) and the merchant which leads to raising both number messages and complexity [127]. To protect the cardholder against spoofing attacks, 3D Secure uses a "Personal-Assurance-Message" (PAM) represented by the issuing bank at time of cardholder authentication [127]. Fig.3 demonstrates "3D-Secure-Verified-by-Visa" Steps. As "Visa-3D-Secure" architecture is demonstrated in Fig.3, twelve messages are exchanged between four parties. We mention its steps briefly: In the first step, the cardholder sends the request to the merchant by using card information. Then the merchant website forwards the cardholder information to the Visa Directory Server (DS). The DS after authentication of the merchant by using either certificate or password contacts the issuer bank regarding to the account number and then forwards the message to the merchant website to confirm the hard holder enrolment. After verification by ACS an acknowledgement along with a URL is transferred to the merchant. The DS sends the data back to the merchant. After, the merchant transfersthepayment authorisation(needed for the ACS) via the cardholder browser. The ACS displays the PAM to assure the cardholder and then request for authentication information from the cardholder. The ACS after receiving and verifying the cardholder information sends its response to the merchant via the cardholder's browser. Finally, the merchant creates the cardholder's receipt and transfers a token to the acquirer bank for settling.

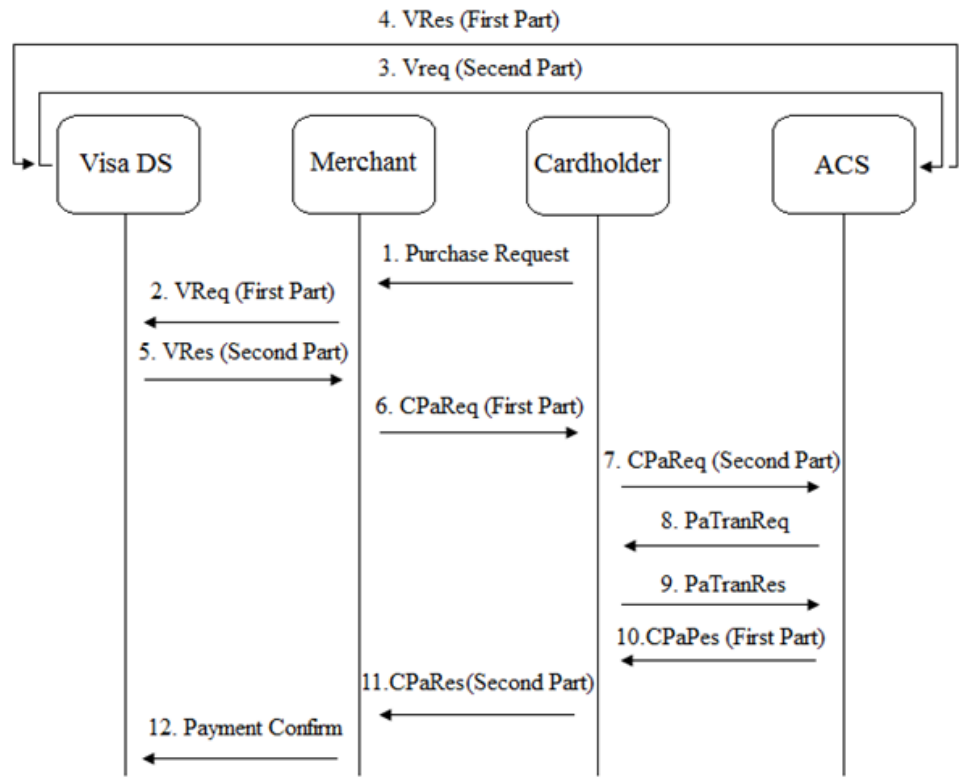

Fig. 3. 3D Secure Verified by Visa Steps (VReq: Verify enrolment Request, VRes: Verify enrolment Response, CPaReq: Condensed PayementAuthorisation Request, PaTranReq: Payment Authorization Transaction Requst, PaTranRes: Payment Authorization Transaction Response, CPaRes: Condensed PayementAuthorisation Response)

*4.1.2. 3D Secure Vulnarebility.* At the begging, 3D Secure using a pop-up form would attempt to get the cardholder's password. But because of the difficulties with the pop-up blockers, it was replaced with inline-frames (abbreviated iFrame) [125]. As a result, because of using an iFrame form without an address bar to receive the cardholder's password, a customer will not be able to verify who is asking for the password. On the other hand, we should consider that most people do not have enough knowledge about "SSL sign", on the web browser.

**4.2  Secure Electronic Transaction (SET)**

The second approach to provide the security of the online payment over the open networks is SET protocol (Secure Electronic Transaction), implemented for personal computers which is as the pint-of-sale. The SET specification has been published publicly for the industry [129].

In practice, SET could not succeed in comparison with 3D Secure. In accordance with [128] the two main reasons for this are related to both the customers and the merchants. The first reason is complexity of SET wallet installation and downloading the certificate for consumers and the second one is expensiveness of both the software and the hardware for the merchants.

To protect the payment process, SET uses symmetric and asymmetric approaches and also key management process is based on Public Key Infrastructure (PKI) [130] [131].

The 3D SET means adapting the SET protocol with the concept of Three Domain (abbreviated 3D) by implementing SET Wallet Server to let the issuer bank to make the transaction on the side of thecustomer which means that the issuing bank becomes the host of "SET-Wallet-Server" [128].

*4.2.1. "SET-Wallet-Server" Solution Steps.* "SET-Wallet-Server" solution' steps are as follow: Selecting the SET as payment approach by the cardholder and then sending a "SET-wake-up" message to the customer by the issuer bank and redirecting this message to the "SET-Wallet-Server" by the cardholder's browser and thereafter, receiving and verifying the cardholder's secret authentication code and finally performing the transaction on behalf of the customer by the issuing bank [128].

*4.2.2. SET Advantages.* In this section we mention the advantages of SET protocol. In accordance with [132] the SET assure the confidentiality of the cardholder's information during the transmission and the storage, by preventing the merchant to be able to see the cardholder's information via encrypting the data using the acquirer's public key and also, assuring the merchant's privacy by preventing the acquirer bank to learn the cardholder's order details which is stored in the merchant's server.

*4.2.3. SET Weaknesses.* Some of main disadvantages of the SET is the cost of implementation that is more expensive and more complicated than SSL/TLS. Also, it is not permitted to place an order to a computer on which the SET requirements has not been installed, because of conducting the SET transaction by cardholder's private key that is stored in the cardholder's SET-initialised computer. Another weakness of SET is related to its performance that may lead to unacceptable transaction speed [132]. In the following sections, we describe EMV based CNP transactions i.e. SET/EMV, EMV/CAP. This type of CNP protocols use two-factor authentication including both EMV smart cards and password.

## 4.3 SET/EMV

One of the solutions to simplify the SET registration is to integrate the SET with EMV to provide more extensive use of SET in internet payments [4]. The other advantage of this approach is enhancing the security by using a two-factor authentication.

*4.3.1. SET and EMV Adoption.* Sharing a number of features by both protocols may lead to possibility of integration. The main purpose of introducing both SET and EMV is reducing the credit cards frauds, although in different types of Card-not-present and Card-present transactions, respectively. On the other hand, both SET and EMV use a hierarchical PKI architecture, such that the credit card brands sign the certificates for the banks and the banks sign the certificates for both the cardholders and the merchants. Another common feature between two protocols is to enter a PIN for the cardholder verification process before performing a transaction [4].

*4.3.2. SET/EMV Analysis.* There are two main problems concerning traditional SET, including the necessity of the customer to register for a certificate to perform any transaction. Also to initiate the SET, the cardholder needs to activate the stored private key at his computer. The other problem is vulnerability and the risk of accessibility of the cardholder's private key at his computer. Whereas, in case of SET/EMV, using an existing key pair and a stored private key in the EMV smart card leads to avoiding the necessity of initial enrollment and also decreasing the probability of revealing the private key [133]. SET/EMV has some advantages for the cardholder: for example, difficulty for stealing or revealing the cardholder's financial information. By using the EMV smart card, the cardholder's private key is protected by the smartcard physical security features. Also, it leads to inability of the merchant to access the payment data, thanks to verification of the merchant by the acquirer bank and signing and encrypting the payment information by using the cardholder's private key and acquirer's public key, respectively [133]. Also, there are some benefits for the merchant: For example, ensuring the cardholder verification by entering the PIN to the EMV smart card via cardholder's computer. On the other hand, the merchant is not reliable against the disputed transaction, thanks to conformity of the SET model with face-to-face traditional transactions (i.e. CP transactions) and also ensuring that nobody is able to access to the cardholder's order information even the acquirer bank which leads to preserving the merchant privacy [133]. In the other side, requiring an additional device that is IC card reader causes that SET/EMV becomes a little complicated for the consumers [143] and also e-consumers prefer not to spend more money to buy an extra device only for performing some e-commerce transactions [144].

### 4.4 Chip Authentication Program: EMV/CAP

The purpose of creating the Chip Authentication Programme (abbreviated EMV/CAP), was to reduce the losses arising from the online payments frauds [134]. The EMV/CAP specifications is not published publicly and it leads to lack of a public security examination. However, a reverse engineering has been done [134] and some useful information is leaked. According to this information, some ambiguities is revealed concerning transactions process and transferring zero amount [135]. EMV/CAP uses a card reader along with a credit Smart Card to create the "one-time codes" for both transaction authentication and login [134].

According to reverse engineering introduced in [134], the EMV/CAP is performed in three modes including: identify (which returns a one-time code), respond (which requires a get change process) and sign (which requires an account number). The "response code" is a MAC calculated over the cardholder's information, plus a "transaction counter" and a flag to show if the PIN is matched. Performing each mode is required entering a PIN by the cardholder. The Fig.4 demonstrates a response transaction [134]. The transaction details which are shown to the user are not signed by the EMV smart card which causes the protocol to be more vulnerable against some type of attacks [137].

The EMV/CAP is similar to EMV traditional transaction. During a transaction, two "Application Cryptograms" are generated by the EMV smart card. The first one is called as Authorization Request Cryptogram (abbreviated ARQC) to authorize the online transaction and the second one is called as "Application Authentication Cryptogram" (abbreviated AAC) to terminate properly the transaction. The cryptograms have not been signed by an asymmetric algorithm, but they are "Message Authentication Codes" which are createdbya symmetric key between the issuer and EMV smart card [135].

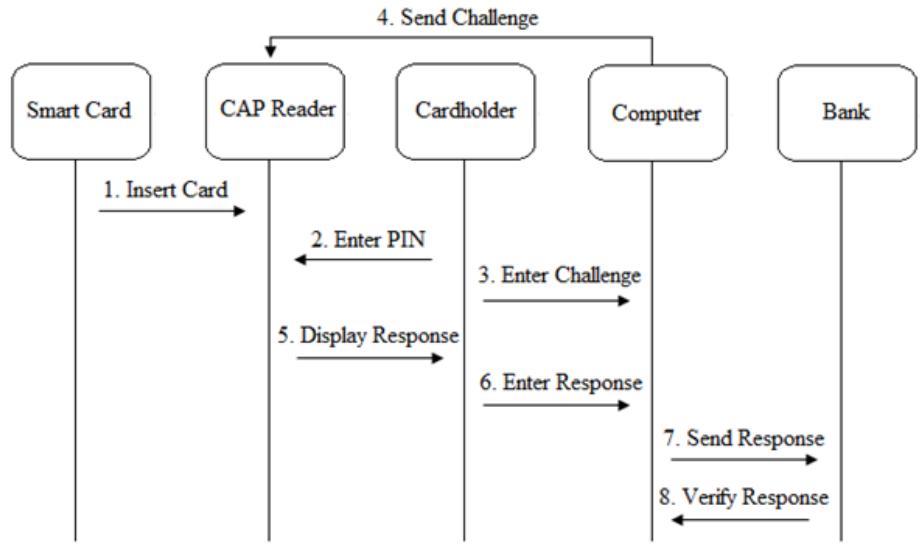

Fig. 4.    EMV/CAP Steps Reverse Engineered by [134]

## 5. TOKENIZATION

The concept of Tokenization was introduced in 2013 by three credit card brands (i.e. "MasterCard", "Visa" and "American Express") to enhance the user's confidentiality and privacy in the online transactions. The concept is to replace the "Primary Account Number" (abbreviated PAN) with a digital value named as the "token" to assure the privacy of the cardholder's sensitive information. The additional steps to the traditional transactions are requesting and providing a token and thereafter, the cardholder will use this token instead the PAN during the transaction. In the other hand, the merchants and the digital wallet operators no longer need to store the PAN as a sensitive data and this adds an additional security layer to the online transactions [14] [15].

The proposed concept need to have the following key features: [15]

•       Providing the more complete information concerning the transaction by using new data fields to improve the fraud detection and accelerate the permission and authorization process.

•       Reliable approaches to verify the cardholder, before replacing the PAN with the token.

•       Designing a standard to achieve the simplification of the transaction process for the merchants in any type of payments such as contactless, online payment etc.

On the other hand, the token generation also needs to have some properties:

•       Ensuring the acceptability of the token largely as a substitute value of the traditional Primary Account Number

- Ability of the all current e-payment participants to perform a transaction using the token.
- Ensuring the feasibility to develop secure innovative e-payment products and applications comfortably.
- Enhancing the cardholder security and confidentiality at time of using the PAN in unreliable environments

## 6. BLIND SIGNATURES SCHEMES IN E-PAYMENT

The "blind signature" and also the "group signature" are the popular approaches for implementation of the e-payment systems [30].

The "Blind signature" proposed by David Chaum [31] is a type of "digital signature" in cryptography in which the message has been changed (or blinded) before signing. The Blind signature is often used in a protocol where the privacy is a major key, such that the signer is able to learn "neither the message nor the resulting signature" and because of this feature it is capable to be used for realizing the protocols which provide the anonymity of participants [31][32].

The key property of a cash is anonymity of the customer, which means if the customer takes some money, the bank has no knowledge about what the customer buys and when spends the money, on the other hands, the merchant has no information about who the customer is [33].

But in online payment, the cardholder has to submit his personal information to the merchant and on the other hand, needs to inform the credit card's company about the merchant. As a result, the potential of the ignoring the cardholder's privacy is higher [33].

In accordance with [33], the blind signature provides the anonymity of the cardholder with following scenario:

The cardholder to withdraw some money from his account creates a coin so-called as $C$ and generates its hash value by using a public hash function called $f(C)$ and then encrypts the hash value by using $E_C(f(C))$. The bank signs encrypted value by using $S_b(E(f(C)))$ and debits the cardholder's bank account. The cardholder then computes $D_C(S_b(E_C(f(C))))$ to decrypt himself encryption to learn $S_b(f(C))$. Then he makes sure that $S_V(S_b(f(C))) = f(C)$ to verify signature of the bank. To spend his electronic cash, he gives both $S_b(f(C))$ (as the bank's signature) and $C$ (i.e. coin) to the merchant and then the merchant sends these to the bank. The bank after checking the validity of the signature, pays the merchant (the cardholder's $C$) and finally marks the coin $C$ as already been spent.

So, in this way, the cardholder's anonymity has been preserved and the bank is able to detect the double-spending, but because of preserving the cardholder's anonymity, cannot punish the double-spender.

Chaum in [34] and [35] has introduced unconditionally untraceable electronic payment using the RSA digital signature scheme. The authors in [36] have used alike idea to provide an electronic-cash protocol which preserves the customer's anonymity until he does not perform a fraudulent transaction and as a penalty the customer's identity will be revealed. In this proposed model, because of using digital signatures, the customer's frame-up is protected only computationally and not unconditionally.

However, the customer's privacy is protected unconditionally. In scheme of [36], the bank does not know where the money has been spent and how much is the individual transaction amount. Authors in [37] introduce a new primitive called as restrictive blind signature to provide untraceable off-line cash in wallets. Authors in [38] present an anonymous payment system with reducing the size of the databases. Authors in [39] presentan off-line electronic check schemes in which the withdrawal and the payment of the check are certainly unlikable, but in the event of double-spending, the customer's ID will be disclosed with high probability. Authors in [40] introduce an untraceable universal electronic cash to solve the problems intrinsic to the real cash in which the customer is able to subdivide his cash balance into some partssuch that the sum of all subdivided parts is equal to his global cash.

A "blind signature"includestwo entities and without a trusted third party: the sender of message to sign along withthe signer which permits the sender receives a signed message without revealing the message and the signature to realize the anonymity of customer (i.e. sender) [41].

Several proposal has been introduced to realize the blind signatures such as [43] [44] [45]. However, all of them provide "full" unlinkability which means impossibility for signer to link a "message-signature pair" to the related signing protocol. This anonymity leads to the possibility of misusing. Typically, in an anonymous e-cash the blind signature causes preventing to link the withdrawal and the payment which is made by the same sender (i.e. customer), a property which leads to possibility of the blackmailing or money laundering [41] [42]. Even in case of small transactions, it would be possible to perform several transactions with the small values and consequently transferring the total value of transactions anonymously. As a result, it could be helpful to remove the anonymity property by using a third trusted party when it is necessary for a legitimate reason to prevent such these abuses [41].

**6.1  Fair Blind Signatures Schemes in E-Payment**

Since the blind signature's properties provide fullunlinkability, it may be helpful widely for untraceable e-cash [46] [37]. However, such these systems might be misused, such as money laundering or blackmailing [41] [47].

In accordance with [41] a novel scheme of "blind signature" has been introduced named "fair blind signature". Such a scheme of "blind signature"includes "several senders" (as the customer), a "signer" (as the bank) and an additional third trusted entity (called as the judge) along with two protocols: a signing protocol and a link-recovery protocol. The additional trusted entity (i.e. judge) delivers the necessary information which allows the signer could be able to derive a link between the "message-signature pair" and the signature protocol if it would be required for a legal reason when the unlinkability property is abused. The signing protocol lets the sender receives a message signed by the signer, such that the signer is not able to link "his view of the protocol" to the "message-signature pair" and the "link-recovery protocol" lets the signer to obtain necessary data via the judge entity. The authors in [41] introduce two different type of fair blind signatures: In the first type, the "link-recovery protocol"permits the signer to determine the related "message-signature pair" and thereby the bank is able to determine the doubtful destination, whereas the second type allows the signer to determine the sender of the message and consequently the bank is able to determine the origin of the doubtful money. Fig.5 demonstrates these two schemes.

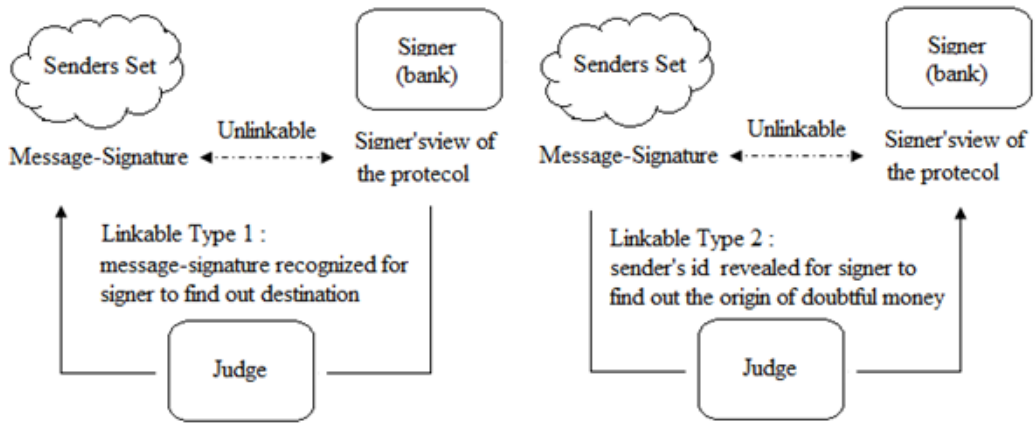

Fig. 5.   Fair blind signature scheme introduced in [41] Type 1 : The "link-recovery protocol" permits the signer to determine the related "message-signature pair" and thereby the bank is able to determine the doubtful destination, Type 2 : The link-recovery protocol allows the signer to determine the sender of the message and consequently the bank is able to determine the origin of the doubtful money.

Thus, by using electronic system on the basis of a "fair blind signature", performing the "anonymous payment" is yet possible, but it is possible to remove the "anonymity" by using a trusted entity which does not need to be involved in the transaction itself [48].

**6.2   Group Blind Signatures Schemes in E-Payment**

The group signature permits signing a message by the members on the side of the group without demonstrating the signer information and the only entity which is able to designate the signer information is manager of group [49].

The concept of the "group signature" is introduced in [50] and improvement of the solution is presented in [51], [52] and [53]. Contrary to all these solutions which have two main unpleasant properties including dependency of the public key size of group on the number of group's member which leads to be problematic in case of large group and also need for modifying the public key in case of adding a new member, in the scheme presented by [49] the group manager is able to add new members without modifying the public key.

In Massachusetts Institute of Technology (MIT) a practical group blind signature has been introduced in [54] which combines existing notions of both "blind signatures" and "group signatures". This model adds the blindness property to the "group signature" scheme which is introduced by Camenisch and Stadler in [49]. In accordance with [54] more than one bank can give out"anonymous" and "untraceable"e-cash securely with remaining the issuer's ID concealed. Also, similar to [49] the "space" and "communication" complexity of the activitiesare not depending on the group size.

However, the classical "group signatures" and "blind signatures" are on the basis of the computation complexity, e.g. "factorization, discrete logarithm" or "quadratic residue problems" which cannot provide unconditional security [30] [55].

**7. QUANTUM CRYPTOGRAPHY IN E-PAYMENT**

In this section, we try to explain the roll of the "Quantum Key Distribution" (abbreviated QKD) in electronic payment systems to achieve unconditional security rather than only computational security which is depends on the attacker's computational power. In this way, we mention some innovations in combination of QKD with electronic payment to learn an advanced security. The first approach is use of QKD in smart card (as a hand-held device) and the second one is a protocol based on QKD to improve the security between two banks.

"Quantum Key Distribution" (abbreviated QKD) is a key distribution approach to generate a secret key over an unconfident channel. Then the secret shared key usable for encryption over any "insecure classical channel". QKD is depended on the quantum physics laws to realize "unconditional security" [16-22].

### 7.1 Quantum Fair Blind Signatures in E-Payment

The "quantum fair blind signature" introduced by [32] includes four steps: "initializing", "signing", "verifying" and "link-recovery" along with three default fair blind signature entities i.e. a "signer", a "trusted entity" and "several senders". In accordance with [32] a "fair blind signature" must include several attributes that are unforgeability, undeniability, verifiability, blindness and traceability. A secure quantum fair blind signature must be able to ensure that nobody can generate a signature except either the signer or the trusted entity. However, authors in [16] show the possibility of the counterfeiting a valid signature by a sender via intercepting a signature in this scheme.

Despite many researches on "quantum signatures", [56-59] the research on use of the QKD in electronic payment is not still satisfactory [30].

One of the attempts in this domain has been introduced in [30] that we describe it in the following section.

The authors in [60] introduce an e-cash protocolon the basis of both techniques of "Quantum blind signature" and "one-time pad" to guaranty the "unconditional security" in electronic payment system. Authors in [61] introduce another e-cash protocol on the basis of"quantum blind signatures" and "group signatures" along with two trusted parties rather than one for enhancing robustness of the system. However, both these systems have a limitation in real life that is possibility of the performing transactions only within the same bank. The Authors in [30] introduce an electronic payment protocol based on "one-time" pad and "quantum proxy blind signature" to solve this problem, allowing supporting the secure transactions between two different banks. The reason for use of proxy signature is that the issuer bank after deducting the money, delegates the blind signature to the merchant's bank [30].

### 7.2 Compact Quantum Key Distribution in E-Payment

Unlike the "classical cryptography" in which the security is relied on the inability to solve a certain mathematical problem in a feasible runtime, the QKD security is based on the quantum physics laws and thereby existence of any eavesdropping and intercepting the quantum communication will be detected by two legitimate parties and as a result, finally a secret key securely will be shared between two parties. The most well-known QKD protocol is the first one called as Bennett-Brassard-84 (BB84) created in 1984 [23]. Here we describe the protocol briefly.

At the beginning, the first party transfers a random sequence of photons which are polarized randomly in four states (including the degrees of 0, 45, 90, 135), without informing the second party. Then these photons are measured randomly by the second party based on the either rectilinear or diagonal without informing the first party after calculation of the quantum transmission, the second party makes public both the received photons and the measurement basis over a classical channel without the actual results. The first party determines the correct measurement basis and then both parties discard the others. If both parties use the degrees of 0 and 45 to demonstrate the value of 0 and the degrees of 90 and 135 to demonstrate the value of 1, as a result, the random string will be the shared key between two parties. If the achieved result by two parties is not the same, it means that there is an eavesdropping over the channel and consequently, they retry to generate another key until achieving a secured shared key without any eavesdropping [24]. Existing an eavesdropper over the channel will be determined by measuring the error rate when exceeding a threshold [25] [24].The idea proposed in [26] is a free space QKD system to use the QKD approach based on BB84 protocol along with use of One-Time Pad (abbreviated OTP) cryptography to communicate a "hand-held electronic device"like a Smartcard or mobile phone with an ATM by using free space optics.

QKD provides a new way for two communication parties to achieve a shared secret key over a long distance. The main advantage is the unconditional security. Since introducing the QKD, it has been progressed quickly in comparison with the original QKD protocol [26][27][23].

The idea proposed in [24] uses the compact QKD approach to protect the interface between the Automated Teller Machine (abbreviated ATM) and the customer's hand-held device (e.g. cardholder's smart card) for achieving the impossibility of any eavesdropping to learn the key information via the skimming attacks called as "false front" on the ATM by which both the key and the card information are revealed [24]. The cardholder's module is combined in a "hand-held" secure device"like a Smartcard and the other module is combined in an ATM. The customer first must register in the issuer bank and after registration, the cardholder receives a smart card which consists of a unique secret string as a One-Time Pad shared with a central server connected to the all ATMs to authenticate and encrypt the further transactions either at the time of withdrawing from an ATM or performing an online transaction. The cardholder must renew periodically this secret shared as OTP [24].

A One-Time Pad (OTP) is a cryptographic protocol in which both parties have an associated random variables, as the "key", which are not related with any variables owned by an eavesdropper [28][29]. In case of use of one-time key variables, the both parties can transfer a secret message over an insure channel [28].

## 8. ELECTRONIC CURRENCY AND PEER TO PEER PAYMENT SYSTEMS

Bitcoin was introduced as a pure peer to peer [112] electronic currency or cryptocurrency [113] to aim at fully decentralizing [114] electronic payment transactions, allowing to perform online payment transactions directly from one party to another one "without" interference of a financial institution as a "trusted third party" [112]. It uses digital signature to verify the bitcoin'spossession and employs Blockchain to stop "double-spending" [116].

Blockchain is broadcasted via a "peer to peer" network to achieve a consensus about the history of transactions via a proof-of-work system [118][119][116] which means need for verification of a "majority of computational power" which is performed by "honest parties"who follow the protocol as described [120]. In fact, the public logs leads to achieving transparency for satisfying the trust of the users, but it has also some negative effects on the privacy and anonymity of the users [121].

Due to publishing transaction logs fully publicly, the user's anonymity is protected only via use of pseudonyms addresses [123].

In accordance with [122] bitcoin provide a weak anonymity due to using pseudonymous addresses which usually causes that the user's transactions may be simply linked together and also in case of linking one user's transactions to his identity, all of user's transactions could be revealed.

The user's privacy in electronic transactions comes back to Chaum's proposal about "anonymous" e-cash by using the "blind signatures" [122]. However, because of the natural property of the traditional electronic payment i.e. need for a central trusted party (commonly a financial institution or bank), the anonymity has not been still seen a "widespread consensus" [122]. According to [123], a main problem of electronic cash protocols is relying on a trusted currency issuer.

One of the main problem of using financial institutions as a trusted third party to perform electronic payments transactions is impossibility of a perfect non-reversible transaction, because financial institutions cannot avoid mediating disputes that is the intrinsic problem of the trust based model and to improve this problem, bitcoin uses cryptographic proof based model trust based [112].

We mention here briefly some most important attempts concerning electronic currency and peer to peer electronic payment systems: Ripple [106] is an electronic currency such that all users are able to issue a currency but it will be accepted only by the peers who has reliance on the issuer and also the acceptance of the transaction needs for a chain of trusted mediators between sender and receiver. i-WAT [107] is another similar system where the chain of trusted mediators is initiated using digital signatures. KARMA [108] was introduced as another electronic currency in which central authority is including a set of users who manage all transactions that in the event of numerous transactions leads to the overhead. PPay [109] is a "peer to peer" network in which the coin issuer has responsibility of keeping track of the transaction. However, it has the same problem of KARMA concerning the overhead. Mondex as smart card electronic currency [110] is similar to an issuer bank for currency issuance that stores the value as electronic information on a smart card instead of the physical notes and coins [111].

Bitcoin is introduced as an "electronic coin" which works as a sequence of "digital signatures" where owners send bitcoin to the succeeding party after adding to the end of bitcoin his signature (generated by his private key) along with the hash of preceding transaction along with public key of the succeeding owner (i.e. receiver) and as a result, the final receiver (i.e. the merchant or the payee) is able to verify the ownership's chain by verifying the signatures [112].

However, the remained problem is inability to determine performing "double-spending" via one of the owners in the chain. To avoid using trusted central authority such as the coin issuer, all transactions must be announced publicly [115] to achieve an agreement on a "unique" history of the transaction by all participants [112]. To avoid double-spending, the final receiver (i.e. merchant) needs to receive the majority of nodes agreed that it was first received [112].

The history of transaction as a public log is called as blockchain which is maintained, recorded and run by a group of minerswho are rewarded regarding to their participation in the bitcoin network [113].

The authors in [113] introduces an attack to demonstrate the bitcoin is not an "incentive-compatible" protocol which means possibility of miners colluding (called as selfish miners) to learn a reward larger than their fair share. Consequently, the "rational"(or selfish) miners would rather cooperate with the "selfish miners". As a result, the size of this colluding group will be increased until converting to a majority. Since, the bitcoinnetwork needs a majority of honest miners who perform exactly the prescribed protocol, if a colluding group is able to become a majority as a result, it can e.g. prevent some transactions [113] or decide to reject originated coins belonging to a particular address leading to devalue of their coins and thereby other users become disinclined to use these coins for the payment [114].

In the other word, the decisions need to be accepted via majority of the computational power that by default is supposed to be honest [114]. In fact, the users vote according to their computing power to prevent double-spending. This action beneficially restricts the power of a specific group of users and leads to performing Sybil attack (i.e. forging different IDs in a "peer to peer"networkto reduce redundancy) [114] [117].

However, [114] demonstrates some critical decisions in the bitcoin's network which may be controlled by a small set of entities, despite controlling enormous computing power in the system. On the other hand, some actions such as protocol updates are not considered to be decentralized, but they are managed by limited entities as administrators without respecting the "computing power" that they control, but depending on their "function" or "duty" in the network. Authors in [114] introduce a suggestions to enhance the decentralization of bitcoin, e.g. reducing the impact of mining pools on the bitcoin system.

In accordance with [122], Mixing services exchange the user's coins arbitrarily with other users' coins after taking them to make unclear the ownership, but without protection from theft by the service (i.e. Mixing services).

The authors in [122] introduce a Mixcoin, fully compatible with bitcoin that adds an accountable service (called as Mixes) to determine the theft of service, means the "accountable" mixes sign warranties to users such that if a user send n coins to Mixes until time of t1, the Mixes have to return n coins to this user by time t2. As a result, in case of misbehaving a Mix, the user can devalue the reputation of the Mix by publishing Mix's warranty.

Another attempt to enhance the anonymity of bitcoin has been introduced in [123] as Zerocoin as a cryptographic extension of bitcoin which despite providing strong anonymity by using unlinking transactions from source of the payment, [124] but an advanced cryptography and considerable changes is needed to be compatible with bitcoin [122]. However, in accordance with [124] it reveals yet the payment destinations along with the transaction amount.

The authors in [124] introduce the concept of Decentralized Anonymous Payment (abbreviated DAP) by which a user is able to pay privately another one, in which the transaction does not reveal the sender of payment, destination and the transaction amount. DAP scheme is used in Zerocash as a decentralized anonymous payment system [124].

## 9. NEAR FIELD COMMUNICATION (NFC)

"Near Field Communication" (abbreviated NFC) was introduced in 2002 for contactless payments [62], as a "short range" and "half-duplex communication", allowing the signal transmission in both directions, but not simultaneously to provide communication between various devices [63]. The communication takes place between two transmitter and receiver devices within few centimetres (about 10 cm (typically less) [68]) and with 13.56 MHz operating frequency [64-66][62]. The more details concerning specifications of NFC is found in ISO 18092 [86].

The NFC devices types is classified as "initiator" to initiate and guide the data exchange and "target" to respond to the request of initiator device. Also, NFC communication consists of two modes: "active" mode in which both "initiator" and "target" use own energy to transmit the data, whereas in "passive" mode the target device uses the energy which is generated by the initiator device [67-68].

In card simulation operating mode, NFC devices employ alike protocols of the Smartcards and are entirely well suited with the standards on the basis of ISO/IEC 14443 including "Type A", "Type B" and "FeliCa" [69].

After the NFC phone is approximated to a reader, the NFC phone treats similar to a Smartcard and the reader communicates with the software of the Secure Element (abbreviated SE) which performs the actions that need high security. The Secure Element is a chipembedded into the phone to provide the security like a Smartcard [70].

### 9.1 NFC Vulnerabilities

Due to the nature of NFC that is a "wireless communication interface", evidently the eavesdropping is one the principal challenges. An attacker will be able trying to extract and decode the data after receiving the Radio Frequency signals either experimentally or by research. According to [68] in case of the data transmitting in active mode, it is feasible to eavesdrop up to 10 m, but in case of passive mode, the distance will be reduced until 1 m. In case of eavesdropping, it is possible also to corrupt the data and disturb the communication such that the receiver cannot detect the attacker sender device [68]. In a DoS attack, the adversary can perform a data corruption by transferring the valid data at a correct time which could be calculated by learning the type of the "modulation" and "coding". However, the attacker cannot modify the actual data. However, NFC phone can detect this type of attack thanks to verify the Radio Frequency field, at time of transferring data. In accordance with [68] the energy consumption of data corruption is notably more than the attack detection. According to [68] data modification is different with data corruption and its feasibility is depended on the amplitude modulation that is used.

The "NFC Data Exchange Format" (abbreviated "NDEF") is used for exchanging data between NFC phones. It is feasible to generate a malicious poster by an attacker with manipulated NDEF tags by modifying a commercial poster [71-73], leading malicious content sharing with the attacker. A solution is signing tags appropriated encryption approaches or using cryptographic tag authentication protocols [72][74].

The other vulnerability of the NFC is Radio Frequency (RF) Interface. In accordance with some studies in [75-78], this type of attack is including "Eavesdropping MITM

Attacks","Data Corruption, Modification and Insertion", "DOS Attacks", "Relay Attacks" and "Phishing by Social Engineering" (a social process by which an attacker learns the information from the user about a targeted attack.) The solution proposed by [68] is to set up a secure communication between NFC phones by using a shared secret via a "key agreement protocol" introduced in [68].

Another type of the NFC vulnerability is related to NFC device and Secure Element, including: running the application on the "Host Controller"in the absence of informing the user, installing the malware on the "Host Controller" or "Secure Element", attack against the NFC controller, adding "modified Secure Element" to the NFC mobile, skimming and token cloning attacks and eavesdropping the "over-the-air" (abbreviated "OTA") [79][80].

"Management of Secure Element" (modifying the data in the SE via the host controller) in which the data stored into the SE may be modified "Host Controller". This property may cause adistant data management the SE and because of this reason, so-called as over-the-air management) [81] [63].

The proposed solutions for such these types of attacks are including: 1. "certificate based authentication mechanisms", 2. "key management policies to authenticate for the secure element measures" [84], 3. "mandatory code signing for NFC API" and 4. "cryptographically linking the application to unique identifiers" [79][80].

### 9.2 Migration from RFID Cards towards NFC Enabled Mobile

The first generation of the contactless payment was developed on RFID credit cards such as PayPass (Mastercard) [145], payWave (VISA) [146] and expressPay (American Express) [147]. In accordance with a research concerning the first generation of RFID credit cards by [84] all the examined cards were susceptible to both cardholder's privacy and relay attacks. These RFID credit cards have two major problems: limitation in application of contactless payment and security problems.

In accordance with [82] the contactless applications are classified as four categories: The first type is Touch and Go where the cardholder only is able to wave the card without any confirmation to perform the transaction and thus only Micro transactions can be performed because of the limitation of the payment's amount (This limitation for MasterCard PayPass and Visa payWave is $50 and for expressPay is $25) [83]. The second type is Touch and Conform in which the cardholder must confirm the transaction, either by using the PIN or other approaches. Thereby in this type, the device must be equipped with an input/output (like keyboard and screen) for confirmation process. The third one is Touch and Connect to perform a "peer to peer datatransferring" and the last one is Touch and Explore in which the cardholder can explore more than one application according to the device capabilities. In accordance with [82] the devices without any energy resource (i.e. battery) are able to perform only the two first types of applications (i.e. Touch and Go and Touch and Conform). On the other hand, because the credit cards have no input/output interaction possibility to confirm the transaction, in case of performing a transaction more than Micro payment, the cardholder has to wave the card a second time to enter the PIN via the POS terminal [82] and thus the RFID credit cards are only able to perform the Touch and Go and Micro payments that because of lack of the interaction confirmation by the cardholder is more vulnerable in security issues. Consequently, RFID credit cards cannot satisfy the contactless authentication benefits i.e. quick and easy [82]. Thus, another device such as mobile phone which is equipped with input/output (i.e. screen and keyboard) and energy resource (i.e. batteries) is more appropriate to perform all four categories of the contactless applications [82].

In this way, there are two type mobile phone capable to contactless or NFC payments: [82] dual chip in which a SIM card is dedicated for the mobile usage along with an NFC chip for the contactless payment and single chip where a payment card is merged to the SIM card as a unique entity for both payment and mobile usage along with an NFC chip for Radio Frequency interaction with the NFC reader [82].

In accordance with [85] the most important security issue in mobile NFC contactless payments is protection of the cardholder (here SIM holder) privacy and also against transaction denial and forgery. In accordance with [82] to achieve this protection level, the main targets are protection of sensitive data, securing both the payment application and the software platform and a tamper resistant hardware.

### 9.3  Google Wallet, Android Pay, Apple Pay

Google Wallet uses NFC along with an android application for mobile payments [87]. Google wallet contains an android application as a user interface to keep safe the wallet by using a PIN, along with several "Java Card applets" installed on the "Secure Element" to store the sensitive data. There is a "Secure Channel Protocol" (abbreviated "SCP02") between the SE and the "remote server" [88].

The android operating system could be the main flaw of the Google Wallet due to its vulnerabilities against different malware attacks [90]. In December 2011 viaForensics (currently NowSecure) divulged several flaws Google Wallet [91].

In another type of attack against Google Wallet, Fuzz testing or fuzzing method which is in fact an approach for discovering the security loopholes in a software, operating system or a network [92] is used as an approach for an attack against Google Wallet in which after feeding the wallet application with corrupted data, the attacker discovers the vulnerabilities to inject designed specific NFC tags to the mobile phone to monitor the results [89][93].

In another attack, an attacker is able to brute force attack against PIN stored on mobile device as an encoded string hash along with visible salt (i.e. a random additional input to a hash function) rather than storing on Secure Element [94-96].

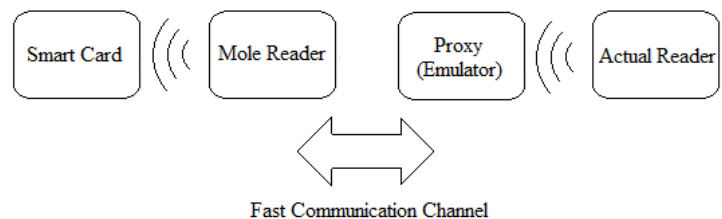

Fig. 6.　　Relay attack on NFC introduced in [88] [97]

A well-known problem with contactless payments is relay attack [88]. A research at Cambridge university [97] demonstrates possibility of a relay attack based on ISO 14443A contactless smart card up to 50 meters in which the attacker uses a contactless Smart Card temporarily in the absence of direct accessibilityand thereby the cardholder will not be able to be informed from the attack. Authors in [98] demonstrates the timing constraints of ISO 14443A leads to limitations to defeat the relay attacks. Also, in accordance with [97] [98] it is not feasible to prevent this type of attacks via application (software) layer.

Fig.5 shows the relay attack requirements [88] [99] including: a relay reader so-called as Mole [97], a card emulator so-called as Proxy [97] to exchange information with the "legitimate reader" and a "fast communication channel".

For the relay attack, in contrary to [97] [138] which use physical layer and bit transferring level, the more recent approaches [139-142] use application layer and transferring Application Protocol Data Units (abbreviated APDU) to relax timing limitations and raising relay distance [88]. One of the limitations in this type of relay attacks (i.e. software based) is that exchanging the APDUs with SE which is a tamper resistant device. The access control schemes of SE are a main flaw which is relied on the "operating system" and thereby "Secure Element" trusts blindly the processor decision which is an insecure element [88].

The second attempt from Google in mobile payment is Android Pay which is a free service allowing the customer to use the mobile which runs android 4.4 or higher for electronic payments at POS terminals [100][101]. It works similarly to Apple Pay [102].

In accordance with the android website's information, the Primary Account Number (PAN) is not sent to the merchant, but it employs a "virtual account number" to show the user information [103].

To support the catch suspicious activities, the customer after transaction, is able to see where the transaction is occurred along with the merchant's name and number [103].

In case of losing or stealing the phone, the cardholder is able to immediately lock the mobile phone from anywhere and renew the password or erasing all personal information [103].

Another similar technology is Apple Pay introduced in October 20, 2014 with iPhone 6 [104].

Apple Pay integrates both biometric authentication and tokenization approach for its mobile payment system [87].

At time of registration, after adding the Primary Account Number (PAN) to apple wallet (called Passbook) a unique and encrypted number (called as Device Account Number) is stored within dedicated SE in iPhone as a token by using Tokenization process. The actual PAN is not stored in apple server and never shared with the merchant or the acquirer bank. The tokenization and de-tokenization process is performed on the card manufacturer network (e.g. VISA or MasterCard) and not in processor or payment gateway [104]. Finally, the mobile sends a one-time transaction code to the merchant [105].

Meantime, Apple Pay needs NFC antenna along with Touch ID on iPhone 6 and a problem of such this payment systems (similar to Google Wallet) is difficulty for a significant number of people to provide them because of their price [87].

## 10. CONCLUSION

In this comprehensive survey, we evaluated all most dominant electronic payment systems and the most successful attack strategies against them, including: Card-present (CP) transactions along with a review of the dominant card-present standard i.e. EMV including several researches to determine several types of attacks against this standard including MITM, pre-play and relay attacks which demonstrates lack of a secure "offline" authentication method that is one of the main purpose of using the smart cards instead of magnetic stripe cards which is participating in authentication process, the evaluation of the EMV migration from RSA cryptosystem to ECC based cryptosystem, the evaluation of the CNP transactions approaches including 3D Secure, 3D SET, SET/EMV and EMV/CAP, the impact of concept of Tokenization and the role of Blind Signatures schemes in electronic cash and E-payment systems, using quantum key distribution (QKD) in electronic payment systems to achieve unconditional security rather than only computational assurance of the security level by using traditional cryptography, the assessment of the electronic currency and peer to peer payment systems such as bitcoin, Near Field Communication (NFC) and contactless payment systems along with the evaluation of the related technologies such as Google wallet, Android Pay and Apple Pay.

This comprehensive survey demonstrates the current status of financial transactions and electronic payment systems. The result demonstrates that in all types of electronic payment systems (i.e. CP, CNP, Contactless, Decentralized) there is some important flaws and weaknesses in security, user privacy, anonymity and performance. It is crucial to enhance and improve the current level of the security of the financial transaction systems, with respect to all necessary properties to achieve a secure and reliable payment system.